\documentclass{article}

\usepackage{natbib}
\usepackage{float}
\usepackage{amsmath}
\usepackage{comment}
\usepackage{fancyhdr}
\usepackage[final]{neurips_2024}
\usepackage[utf8]{inputenc} 
\usepackage[T1]{fontenc}    
\usepackage{hyperref}       
\usepackage{url}            
\usepackage{booktabs}       
\usepackage{amsfonts}       
\usepackage{nicefrac}       
\usepackage{setspace}
\usepackage{microtype}      
\usepackage{xcolor}         
\usepackage{graphicx}

\title{\LARGE\bfseries NoLBERT: A No Lookahead(back) Foundational Language Model}

\author{
  \begin{tabular}{c}
    { \bf Ali Kakhbod,  Peiyao Li} \\
    {\small Haas School of Business}\\
    \multicolumn{1}{c}{\small University of California, Berkeley} \\
    \texttt{\small \{akakhbod,ojhfklsjhl\}@berkeley.edu} 
  \end{tabular}
}

\begin{document}

\maketitle

\begin{abstract}
We present NoLBERT, a lightweight, timestamped foundational language model for empirical research---particularly for forecasting in economics, finance, and the social sciences. By pretraining exclusively on text from 1976 to 1995, NoLBERT avoids both {\it lookback} and {\it lookahead} biases (information leakage) that can undermine econometric inference. It exceeds domain-specific baselines on NLP benchmarks while maintaining temporal consistency. Applied to patent texts, NoLBERT enables the construction of firm-level innovation networks and shows that gains in innovation centrality predict higher long-run profit growth.

\end{abstract}
\onehalfspacing
\section{Introduction}
Text-based methods have become increasingly central to empirical finance (e.g., \citet{EisfeldtSchubert:25}). Yet most existing language models may not be suitable for prediction problems: they are trained on corpora spanning centuries, which can introduce two fundamental (information leakage) biases. \textit{Lookahead bias} contaminates backtests when models implicitly learn from future information, while \textit{lookback bias} conflates meanings across eras, generating temporally inconsistent representations.

We introduce NoLBERT, a timestamped, domain-ready encoder designed for textual inference.\footnote{See the Hugging Face model card: \href{https://huggingface.co/alikLab/NoLBERT}{https://huggingface.co/alikLab/NoLBERT}} Built on the DeBERTa v3 architecture introduced in \citet{he2020deberta}, NoLBERT is pre-trained exclusively on 1976–1995 data, with validation on 1996. This narrow training horizon eliminates both {\it lookback} and {\it lookahead} biases while keeping the model compact (109M parameters). NoLBERT achieves higher performance on GLUE tasks relative to domain models such as FinBERT, while offering temporal constraints helpful for empirical research. Nonetheless, the use of NoLBERT versus industrial-grade large models should be analyzed on a case-by-case basis (see Appendix \ref{app:small_large_comparison} for a detailed discussion of the suitability of each class of models). 

To illustrate its applicability in economic research, we apply NoLBERT to the study of firm innovation and performance. We fine-tune the model on {\it patent texts} to construct firm-level innovation networks, compute centrality measures, and link these to firm profit growth. Our econometric analysis shows that increases in innovation centrality are significantly associated with higher medium- and long-run profit growth. Together, these results highlight NoLBERT’s potential as a foundational tool for economic research: it combines strong language modeling performance with the temporal discipline helpful for econometric inference.

\section{Pre-training NoLBERT}

\subsection{Base architecture and pre-training data}
For economists, a language encoder should (i) produce embeddings that are informative for downstream econometrics and (ii) be portable enough to run at scale on modest GPUs. We therefore adopt \emph{DeBERTa v3 base} and pre-train it on text from 1976--1995, balancing compact size with strong language-modeling performance.

Our pre-training corpus is curated to satisfy this time window while covering diverse, timestamped domains. We combine (a) popular-culture sources (movie scripts, TV dialogues, magazines, novels, blogs), (b) formal prose (parliamentary debates, campaign materials, news, academic papers), and (c) economics-relevant materials (FOMC transcripts, patents). See Appendix \ref{app:data_processing} for dataset construction details.

\subsection{Pre-training procedure and benchmark performance}

We pre-train our base model using data from 1976 to 1995 and use 1996 as validation data to track performance. To strictly prevent temporal biases, we first train a custom ByteLevelBPE tokenizer from scratch with a vocabulary size of 30,000 tokens and a minimum frequency threshold of 2, incorporating standard special tokens for masked language modeling. We then train our model with mixed precision over 15 epochs.

We evaluate the performance of our model on the GLUE benchmark, including CoLA, SST2, QQP, MNLI, and QNLI. As shown in Table \ref{tab:glue}, NoLBERT achieves the highest average performance when compared with similar models, FinBERT and StoriesLM, and it dominates the other two models in tasks other than CoLA (\citet{araci2019finbert,sarkar2025economic}).

\begin{table}[h]
  \caption{Performance comparison of financial language models across benchmark tasks.}
  \label{tab:glue}
  \centering
  \begin{tabular}{lrrrrrrr}
    \toprule
    Model & Vocabulary (K) &  \#Params (M) & CoLA & SST2 & QQP & MNLI & QNLI \\
    \midrule
    FinBERT & 30 & 110  & 0.29 & 0.89 & 0.87 & 0.79 & 0.86 \\
    StoriesLM & 30 & 110  & \textbf{0.47} & 0.90 & 0.87 & 0.80 & 0.87 \\
    NoLBERT & 30 & 109  & 0.43 & \textbf{0.91} & \textbf{0.91} & \textbf{0.82} & \textbf{0.89} \\
    \bottomrule
  \end{tabular}
\end{table}
\section{Lookback and lookahead biases}


\subsection{Avoiding lookahead and lookback biases using timestamped pre-training}
\textit{Lookahead bias} occurs when models trained on future information contaminate inferences. For instance, when predicting stock returns from news articles, a model trained on data including future outcomes may simply retrieve memorized return patterns rather than genuinely inferring from text content. This renders predictions invalid for periods beyond the training data. 

\textit{Lookback bias} arises when models trained across long time periods produce representations that conflate different historical contexts. For example, the phrase "running a program" meant organizing an event in the early 1900s, but executing computer code by the late 20th century. Models trained on century-spanning data create ambiguous representations that may mischaracterize texts from specific periods. We avoid both biases by restricting training to the narrow 1976-1995 window, with validation strictly from 1996, ensuring temporal consistency in our text representations.

\subsection{Validity check}
We empirically evaluate whether NoLBERT’s knowledge is temporally bounded to its pre-training window (1976–1995). To do so, we design two paired $t$-tests: one for \emph{lookahead bias} and another for \emph{lookback bias}. 

For lookahead bias, we construct 20 test words whose meanings shifted between 1976–1995 (old era) and 2020–present (new era). Using GPT-5, we generate sentence pairs in which the focal word is masked—one reflecting the old sense and the other the new. For example, the word  “{\it token}” appears as “{\it He offered a <mask> of gratitude in the form of flowers.}” (old) versus “{\it She invested in a governance <mask> for the new DAO.}” (new). If the model is temporally consistent, it should predict the masked word more accurately in the context of its training era. For lookback bias, we repeat the procedure with another 20 words, where the ``new’’ era is 1976–1995 and the ``old’’ era is the 19th century. 

Let $P_{\text{old}}(w_i)$ and $P_{\text{new}}(w_i)$ denote the probabilities of predicting the masked word $w_i$ in old versus new contexts. We conduct one-sided paired $t$-tests on
\[
\frac{1}{n}\sum_{i=1}^n \big[\log P_{\text{old}}(w_i) - \log P_{\text{new}}(w_i)\big]=0,
\]
with the alternative hypothesis positive for lookahead (favoring old meanings) and negative for lookback (favoring newer meanings).

\begin{table}[h]
  \caption{Validity checks against lookahead and lookback biases.}
  \label{tab:biases}
  \centering
  \begin{tabular}{lrrrr}
    \toprule
    New era & Old era & Difference & $t$-stat & $p$-value \\
    \midrule
    2020–present & 1976–1995 & 2.78 & 5.56 & $<0.01$ \\
    1976–1995 & 19th century & -2.53 & -4.60 & $<0.01$ \\
    \bottomrule
  \end{tabular}
\end{table}

As shown in Table \ref{tab:biases}, the tests strongly reject the null of equal predictive accuracy across eras, confirming that NoLBERT is temporally localized.

\section{Applications}

In this section, we apply {NoLBERT} to compute firms’ innovation centrality scores based on patent texts, and examine their association with corresponding firms' profit growth. 



\subsection{Fine-tuning NoLBERT via contrastive learning and patent text data}
We adopt a bottom-up approach to estimate firm-pair innovation similarities by fine-tuning NoLBERT with a contrastive learning objective. Our aim is to ensure that the document-level ``[CLS]'' embedding captures information relevant to innovation-specific textual similarity.

More specifically, for each patent, we randomly split its text into two chunks, denoted $A$ and $B$. We then construct a balanced dataset as follows: in 50\% of the cases, chunk $A$ is paired with chunk $B$ from the same patent (labeled 1, representing 100\% similarity), while in the remaining 50\% of the cases, chunk $A$ is paired with a chunk from a different patent (labeled 0). NoLBERT is fine-tuned to predict whether a given pair of chunks originates from the same patent.  

We repeat this procedure on an annual basis, fine-tuning NoLBERT separately for patents granted in each year. For each year, we randomly split the data into training and test sets in a 70/30 ratio and train the model for one epoch. Across years, the classifier achieves an average accuracy of $98\%$, indicating that the fine-tuned ``[CLS]'' embeddings provide a strong representation of patent-level similarity.  

Then, we use the trained model in year $t$ to construct the patent-level embeddings in year $t$. Let $S^t_A$ and $S^t_B$ be the sets of NoLBERT-based ``[CLS]'' embeddings of all patents granted to firm A and B in year $t$, the similarity between firms A and B in year $t$ is computed as 
\begin{align}
    \text{sim}(A,B,t)=\cos(\bar{s}_A^t,\bar{s}_B^t),
\end{align}
\noindent
where $\bar{s}_A^t$ and $\bar{s}_B^t$ are the average (pooled) embeddings of $S^t_A$ and $S^t_B$.

\subsection{Innovation centrality and firm growth}
\label{subsec:centrality}
For each year $t$, we first construct a sparse, weighted firm–firm innovation similarity network $G_t$ whose nodes are firms and whose (undirected) edges carry weights equal to the pairwise similarity. This graph is represented by a sparse adjacency matrix $A_t\in\mathbb{R}^{n\times n}$ where the $(i,j)$th entry captures the similarity between firms $i$ and $j$. We then form the row-stochastic transition matrix $P_t = D_t^{-1} A_t$, where $D_t=\mathrm{diag}(d_1,\ldots,d_n)$ with $d_i=\sum_j A_{t,ij}$. Finally, firms' PageRank centralities are computed by power iteration with damping: starting from $p^{(0)}=\tfrac{1}{n}\mathbf{1}$, we iterate
\[
p^{(k+1)} \;=\; \alpha\,P_t^\top p^{(k)} \;+\; (1-\alpha)\tfrac{1}{n}\mathbf{1},
\]
with damping factor $\alpha=0.85$, until $\lVert p^{(k+1)}-p^{(k)}\rVert_1<\texttt{tol}$ or a maximum of \texttt{max\_iter} iterations is reached. We show the summary statistics in Appendix \ref{app:summary}. 

Next, we merge firms' centrality scores with financial and operational characteristics from COMPUSTAT, including profit, capital stock, total assets, and employment. We further incorporate innovation value data from \citet{kogan2017technological}, which allows us to estimate the aggregate innovation value of each firm and industry in each year.\footnote{Details on data processing and merging are provided in Appendix \ref{app:data_processing}.}  

We then estimate the following regressions to examine the relationship between changes in firms’ innovation centrality and their subsequent profit growth:
\begin{align}
\label{eq:main_reg}
    \Pi_{f,t+k} - \Pi_{f,t} \;=\; \alpha^k_1 \,\Delta \text{Centrality}^{PR}_{f,t} \;+\; \beta^k Z_{f,t} \;+\; \delta^k_{t} \;+\; \delta^k_{\text{ind}} \;+\; \epsilon^k_{f,t},
\end{align}
\noindent
where $\Pi_{f,t}$ denotes firm $f$'s log profit in year $t$, and $k \in \{1,2,3,4,5\}$ indexes the forecast horizon. The main independent variable $\Delta \text{Centrality}^{PR}_{f,t}$ is the one-year log change in PageRank centrality, defined as 
\[
\Delta \text{Centrality}^{PR}_{f,t} = \log\!\left(\text{PR}_{f,t}\right) - \log\!\left(\text{PR}_{f,t-1}\right).
\] 
The vector $Z_{f,t}$ includes controls for firm-level innovation value, industry-level innovation value, and the logs of profit, employment, and capital stock. $\delta^k_t$ and $\delta^k_{\text{ind}}$ denote year and Fama–French 30 industry fixed effects, respectively. Standard errors are double-clustered at the industry and year level. All independent variables are standardized within industry-year cells.  

\begin{table}[h] \centering
  \caption{Association between changes in innovation centrality and profit growth.} 
  \label{tab:profit_growth} 
\begin{tabular}{@{\extracolsep{5pt}}lrrrrr} 
\\[-1.8ex]\hline 
\hline \\[-1.8ex] 
 & \multicolumn{5}{c}{\textit{Dependent variable:} $\Pi_{t+k} - \Pi_{t}$} \\ 
\cline{2-6} 
\\[-1.8ex] & $k=1$ & $k=2$ & $k=3$ & $k=4$ & $k=5$ \\ 
\hline \\[-1.8ex] 
 $\Delta \text{Centrality}^{PR}_{f,t}$ & 0.001 & 0.005$^{*}$ & 0.007$^{**}$ & 0.007$^{*}$ & 0.003$^{**}$ \\ 
  & (0.004) & (0.003) & (0.003) & (0.004) & (0.001)\\ 
\hline \\[-1.8ex] 
Observations & 20,976 & 18,987 & 17,204 & 15,615 & 14,199 \\ 
R$^{2}$       & 0.023  & 0.028  & 0.030  & 0.032  & 0.035  \\ 
\hline 
\hline \\[-1.8ex] 
\textit{Note:}  & \multicolumn{5}{r}{$^{*}$p$<$0.1; $^{**}$p$<$0.05; $^{***}$p$<$0.01. } \\ 
\end{tabular} 
\end{table} 

As shown in Table~\ref{tab:profit_growth}, increases in innovation centrality are significantly and positively associated with profit growth in the medium to long run (years $t+2$ through $t+5$). Quantitatively, a one standard deviation increase in PageRank centrality growth is associated with a 0.5\% increase in profit growth by year~2 and a 0.3\% increase by year~5. For robustness, we confirm that using two-year changes in centrality yields similar positive associations with profit growth at horizons $t+3$ to $t+5$. Moreover, the results remain robust when we replace PageRank with weighted degree centrality as the centrality measure.\footnote{Details about these robustness analyses are discussed in Appendix \ref{app:robust}.}


\clearpage
\bibliographystyle{plainnat} 
\bibliography{bib}
\clearpage
\appendix
\renewcommand{\thetable}{A\arabic{table}}
\renewcommand{\thefigure}{A\arabic{figure}}
\setcounter{table}{0}
\setcounter{figure}{0}
\section{Appendix / supplemental material}
In this Appendix, we provide additional details and background information related to our paper. First, in Appendix \ref{app:small_large_comparison}, we discuss the advantages of using small bias-free models and large models with lookahead (back) biases.
In Appendix \ref{app:robust}, we show results of some robustness analyses verifying the significantly positive association between innovation centrality and profit growth. In Appendix \ref{app:data_processing}, we show more details about our data processing steps and summary statistics. In Appendix \ref{app:example}, we show more direct examples to show that NoLBERT is free from lookahead and lookback biases. Lastly, in Appendix \ref{app:kpss}, we show more details about how innovation value and centralities can be interpreted.
\subsection{A bias-performance tradeoff between NoLBERT and large industrial models}
\label{app:small_large_comparison}
Even though NoLBERT has the advantage of no lookahead and lookback bias, researchers should carefully consider their model choice on a case-by-case basis, especially for long texts. 

In particular, there is a bias–performance trade-off between NoLBERT or other custom small models (or simpler NLP methods, e.g., BoW, Word2Vec, etc.) versus large industrial-grade language models. On one hand, a BERT-like custom information-leakage-free model avoids temporal inconsistencies by design. On the other hand, these models lack the ability to process long texts due to limited context windows, and their output text representations are often of lower quality compared to large models trained on unconstrained data.


The advantage of avoiding temporal biases is pronounced in tasks where models must predict outcomes that go beyond the information explicitly stated in the text, such as forecasting stock price reactions from earnings call transcripts, despite the tradeoff of having less precise text representations.
However, for in-context information retrieval tasks such as summarization, classification, and other NLP tasks based on given precise guidelines, the risk of information leakage from the model’s out-of-context knowledge base is limited (with careful prompting and verification, or by using methods like RAG). Therefore, large, highly performant models may be preferable.

\subsection{Robustness analysis of the association between innovation centrality and profit growth}
\label{app:robust}

In this Appendix, we conduct robustness checks of our finding that growth in the focal firm's innovation centrality is significantly positively associated with its profit growth. In particular, we do two types of robustness checks. First, we extend subsection \ref{subsec:centrality} by examining the association between the 2-year (instead of 1-year) growth of centrality and profit growth. Secondly, we use weighted-degree as an alternative definition of centrality and show that growth in innovation centrality is significantly positively associated with profit growth in the medium to long run. Let $F_t$ be all of the firms with at least 1 granted patent in year $t$, and take any focal firm $f$, the weighted-degree centrality of $f$ in year $t$ is
$$\text{centrality}^{wd}_{f,t}=\frac{\sum_{s\in F_t\setminus f}\text{sim}(f,s,t)}{\max_{f\in F_t}\sum_{s\in F_t\setminus f}\text{sim}(f,s,t) }.$$
We are following the specification of equation \ref{eq:main_reg} where the growth in centrality is either 1 or 2 years, and the centrality is computed by either the PageRank or weighted-degree definition. Note that PageRank and weighted-degree centralities are highly correlated with a Pearson correlation of $0.81$.

\begin{table}[h] \centering
  \caption{Association between changes in innovation centrality and profit growth (different definitions of innovation centrality).} 
  \label{tab:profit_growth_robust} 
\begin{tabular}{@{\extracolsep{5pt}}lrrrrr} 
\\[-1.8ex]\hline 
\hline \\[-1.8ex] 
 & \multicolumn{5}{c}{\textit{Dependent variable:} $\Pi_{t+k}-\Pi_t$} \\ 
\cline{2-6} 
\\[-1.8ex] & $k=1$ & $k=2$ & $k=3$ & $k=4$ & $k=5$ \\ 
\hline \\[-1.8ex] 
  $\Delta \text{Centrality}^{PR}_{f,t}$ (1-year)  & 0.001 & $0.005^{*}$ & $0.007^{**}$ & $0.007^{*}$ & $0.003^{**}$ \\ 
  & (0.004) & (0.003) & (0.003) & (0.004) & (0.001)\\ 
  $\Delta \text{Centrality}^{PR}_{f,t}$ (2-year)  & 0.002 & 0.004 & $0.008^{**}$ & $0.010^{***}$ & $0.008^{***}$ \\ 
  & (0.002) & (0.003) & (0.003) & (0.002) & (0.002)\\ 
  $\Delta \text{Centrality}^{wd}_{f,t}$ (1-year)  & 0.001 & $0.005^{*}$ & $0.007^{**}$ & $0.007^{*}$ & $0.003^{**}$ \\ 
  & (0.004) & (0.003) & (0.003) & (0.004) & (0.001)\\ 
  $\Delta \text{Centrality}^{wd}_{f,t}$ (2-year)   & 0.002 & 0.004 & $0.008^{**}$ & $0.010^{***}$ & $0.008^{***}$ \\ 
  & (0.002) & (0.003) & (0.003) & (0.002) & (0.002)\\ 
\hline \\[-1.8ex] 
Observations 1y & \multicolumn{1}{c}{20,976} & \multicolumn{1}{c}{18,987} & \multicolumn{1}{c}{17,204} & \multicolumn{1}{c}{15,615} & \multicolumn{1}{c}{14,199} \\ 
Observations 2y & \multicolumn{1}{c}{20,272} & \multicolumn{1}{c}{18,354} & \multicolumn{1}{c}{16,646} & \multicolumn{1}{c}{15,148} & \multicolumn{1}{c}{13,753} \\ 
R$^{2}$: PR 1y & \multicolumn{1}{c}{0.023} & \multicolumn{1}{c}{0.028} & \multicolumn{1}{c}{0.030} & \multicolumn{1}{c}{0.032} & \multicolumn{1}{c}{0.035} \\ 
R$^{2}$: PR 2y & \multicolumn{1}{c}{0.023} & \multicolumn{1}{c}{0.029} & \multicolumn{1}{c}{0.032} & \multicolumn{1}{c}{0.033} & \multicolumn{1}{c}{0.036} \\ 
R$^{2}$: WD 1y & \multicolumn{1}{c}{0.023} & \multicolumn{1}{c}{0.028} & \multicolumn{1}{c}{0.030} & \multicolumn{1}{c}{0.032} & \multicolumn{1}{c}{0.035} \\ 
R$^{2}$: WD 2y & \multicolumn{1}{c}{0.023} & \multicolumn{1}{c}{0.029} & \multicolumn{1}{c}{0.032} & \multicolumn{1}{c}{0.033} & \multicolumn{1}{c}{0.036} \\ 
\hline 
\hline \\[-1.8ex] 
\textit{Note:}  & \multicolumn{5}{r}{$^{*}$p$<$0.1; $^{**}$p$<$0.05; $^{***}$p$<$0.01. } \\ 
\end{tabular} 
\end{table}

As shown in Table \ref{tab:profit_growth_robust}, across all definitions of growth in innovation centrality (PageRank, weighted-degree, 1-year growth, and 2-year growth), the growth in the focal firm's innovation centrality is significantly positively associated with its profit growth. This shows the robustness of our finding in subsection \ref{subsec:centrality}.

\subsection{Data processing details and summary statistics}
\label{app:data_processing}

\subsubsection{Data processing for pre-training}
We highlight the detailed procedure that we use to process a few key datasets that are included in our pre-training data.

\noindent
\textbf{Processing FOMC minutes}
For each meeting record, we extracted the date and the full text of the minutes, excluded rows with missing or empty text, and split the remaining text into individual sentences. Documents containing fewer than ten sentences were retained in their entirety, while longer documents were partitioned into consecutive chunks of three to seven sentences (the number of sentences is randomized), ensuring variability in chunk size but preserving sentence order. Any residual sentences at the end of a document were grouped together as a final chunk. 

\noindent
\textbf{Processing patent texts}
We include the abstracts of all utility patents in the USPTO in each year. The only filter we apply is removing any abstract longer than 300 words, which eliminates $0.4\%$ of patents.

Other sources of long documents are processed with a similar procedure.

\subsubsection{Data processing for fine-tuning}
\label{app:fine_tuning_data}
To prepare patent abstracts for a similarity-based fine-tuning task, we first filtered the dataset to retain only those entries with more than two sentences. For each abstract, we randomly selected a \emph{breaking point} between the first and the penultimate sentence, and split the abstract into two parts: chunk A (the leading segment up to the breaking point) and chunk B (the trailing segment thereafter). This procedure ensured that each abstract was decomposed into a coherent pair of textual fragments.  

Next, we constructed training data by year of grant between 1997 and 2021. Within each year, abstracts were randomly shuffled, and two versions of chunk B were generated: the original trailing segment and a shifted version, where chunk B was rotated across the set of abstracts. For each abstract, we randomly chose with probability 0.5 whether to keep the true chunk B (positive pair) or replace it with the shifted chunk B (negative pair). A binary indicator variable (\texttt{sim}) was created to mark whether the final pair represented a true continuation of the abstract or an artificially mismatched fragment.  

Then, we fine-tune a NoLBERT-based text–pair classifier on the processed patent data in a year-by-year fashion. For each grant year $y \in \{1997,\dots,2021\}$, we create stratified splits (70\% train, 30\% test). Then, we use the custom NoLBERT tokenizer to process the pairs, padding/truncating each to 512 tokens. 

The classifier instantiates the NoLBERT encoder and extracts the [CLS] embeddings for chunk A and chunk B. We compose a richer pair representation by concatenating four components: (i-ii) direct concatenation $[h_A;h_B]$, (iii) element-wise product $h_A \odot h_B$, and (iv) absolute difference $|h_A-h_B|$, yielding a $4d$-dimensional vector for hidden size $d$. An MLP head (two \texttt{ReLU} layers with dropout) maps this feature to logits over two classes. We optimize with AdamW (learning rate $2{\times}10^{-5}$, weight decay 0.01), a linear warmup/decay schedule (10\% warmup), and cross-entropy loss, applying gradient clipping ($\lVert g\rVert_2 \le 1$) for stability. Training proceeds for one epoch per year, with scheduler steps taken per batch. 

\subsubsection{Creating firm dataset}

In our econometric analyses of innovation centrality versus profit growth, we use five firm characteristics from COMPUSTAT: ``Property, Plant, and Equipment – Total'' (ppegt), ``Assets - Total'' (at), ``Sales'' (sale), ``Cost of Goods Sold'' (cogs), ``Employment'' (emp). In particular, a firm's capital stock is computed as ppegt divided by the equipment deflator of each year. Profit is computed as sale-cogs divided by the consumer price index of each year. 

In addition, we use year and PERMNO to merge the COMPUSTAT data with the innovation value dataset from \citet{kogan2017technological}. A firm's innovation value is computed as the sum of patent values granted in each year divided by its total assets. The value of a patent is estimated as the real stock market reaction to the application and granting of the patent. The industry-level innovation value that each firm is exposed to is the aggregated firm-level innovation value in the focal 3-digit SIC industry, other than the focal firm.

\subsection{Summary statistics}
\subsubsection{Number of samples from each data source}

\begin{table}[H]
\centering
\caption{Summary statistics of text sources.}
\label{tab:summary_stats}
\begin{tabular}{l r}
\toprule
\textbf{Source} & \textbf{Count} \\
\midrule
Paper        & 2,815,934 \\
COCA         & 1,566,883 \\
Patent       & 1,102,398 \\
Song         &   284,506 \\
Book         &   140,810 \\
Parliament   &    76,470 \\
Hong Kong News     &    32,914 \\
Reuters News       &    24,067 \\
Music        &     1,149 \\
FOMC         &       678 \\
Political Campaign     &       615 \\
Friends (TV Show)     &       198 \\
\bottomrule
\end{tabular}
\end{table}

\subsubsection{Summary statistics of PageRank centrality}
\label{app:summary}
\begin{figure}[H]
    \centering
        \caption{Summary statistics of PageRank and weighted-degree centrality.}
\label{fig:placeholder}
\includegraphics[width=0.8\linewidth]{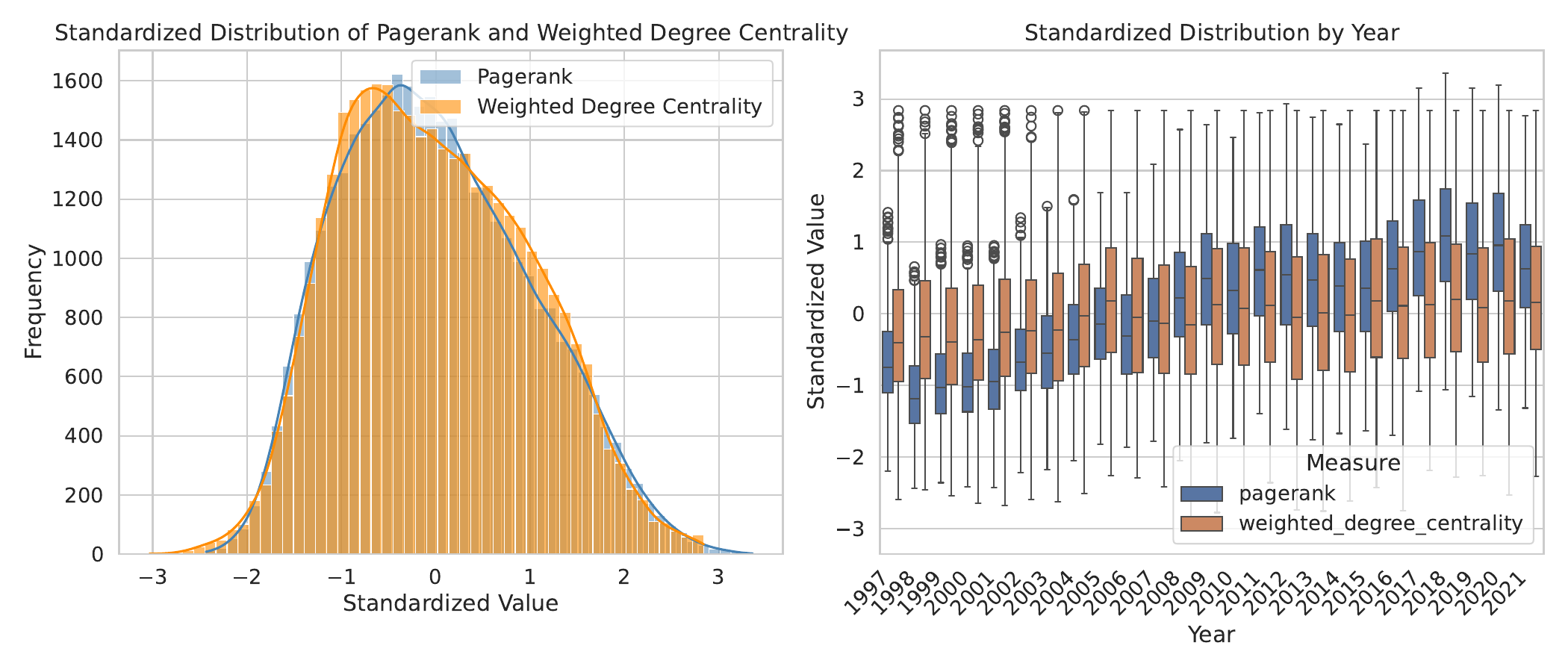}

\end{figure}

\subsection{Demonstration of NoLBERT against lookback and lookahead biases}
\label{app:example}

We demonstrate the knowledge limitations of the NoLBERT model by asking it to fill in the job title of United States presidents from 1969 to 2008. The prompt is
\begin{quote}
    XXX is a United States <mask>.
\end{quote}

As shown by the results in Table \ref{tab:example}, other than George W. Bush, the probabilities of all presidents outside of our training interval (1976-1995) are strictly lower than any president in our time interval. In addition, all of these presidents within our time frame are correctly identified as a president within the top 2 most likely words. Presidents who took office outside of the time range are not identified within the top 2, other than George W. Bush. In the case of George W. Bush, we find that the last name Bush implies presidency in the knowledge base of NoLBERT (because it has seen George H.W. Bush as a president in the training data). Indeed, when we only mention ``Bush'' and ``George Bush'', NoLBERT also predicts this person as the United States president. Overall, this provides more examples to demonstrate that NoLBERT's knowledge is restricted within the time frame from 1976 to 1995.

\begin{table}[H]
\centering
\caption{Examples demonstrating no lookahead and lookback biases.}
\label{tab:example}
\begin{tabular}{l r r r}
\toprule
\textbf{Name} & &\textbf{Within Top 2 Choices} & \textbf{Probability} \\
\midrule
Richard Nixon     & & No  & 0.05 \\
Gerald Ford        && No  & 0.03 \\
\midrule
&\textit{1977}      &     &      \\
Jimmy Carter       && Yes & 0.14 \\
Ronald Reagan      && Yes & 0.58 \\
George H.W. Bush   && Yes & 0.51 \\
Bill Clinton       && Yes & 0.54 \\
&\textit{2001}      &     &      \\
\midrule
George W. Bush     && Yes & 0.45 \\
Barack Obama       && No  & 0.08 \\
\bottomrule
\end{tabular}
\end{table}

\subsection{More details about innovation value and centrality}
\label{app:kpss}
In this subsubsection, we unpack our PageRank innovation centrality measure to show how its distribution evolves over time and how it corresponds with other firm characteristics. In addition, we demonstrate one potential mechanism through which growth in centrality is associated with the focal firm's profit growth.
\subsubsection{Industry concentration among the most central firms}

\begin{figure}[H]
    \centering
        \caption{Cumulative industry composition from $t$ to $2021$ of the most central firms.}
    \label{fig:ind_concentration}
    \includegraphics[width=\linewidth]{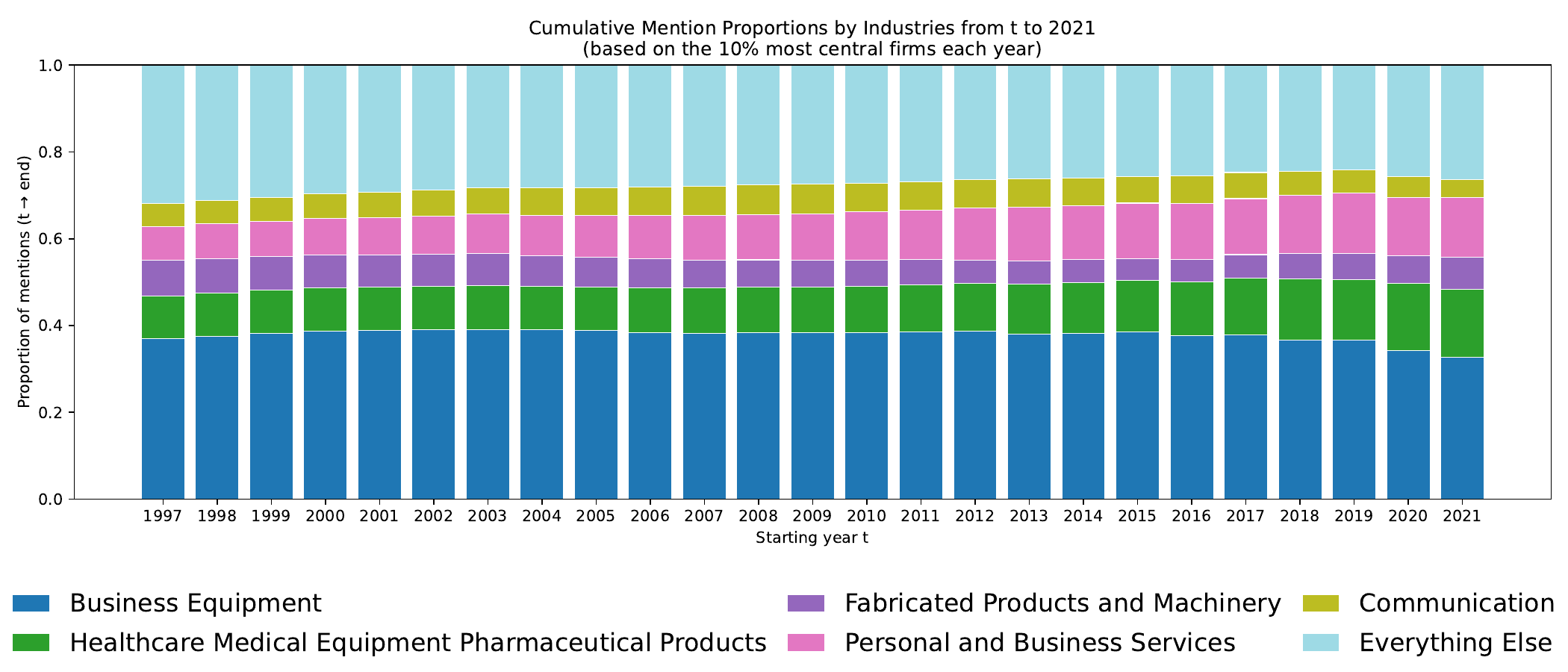}

\end{figure}

Figure \ref{fig:ind_concentration} illustrates the industries that account for the largest shares of the most central firms (defined as the top 10\% in each year). Specifically, for each year $t$, we identify the most central firms, record their industries, and then plot the cumulative industry composition from year $t$ through the end of our sample in 2021.  

Over time, two industries increasingly dominate the set of most central firms: \emph{Personal and Business Services} (e.g., Apple, Alphabet) and \emph{Healthcare, Medical Equipment, and Pharmaceuticals} (e.g., Johnson \& Johnson, Pfizer). Correspondingly, the residual ``Everything else'' category shrinks. This pattern highlights a growing concentration of innovation centrality within a small set of industries, particularly those in technology-related services and healthcare.

\subsubsection{Innovation centrality and other firm characteristics}
\label{app:centrality_characteristics}
We compute the Pearson correlations between our innovation centrality measure and a set of firm characteristics (log profit, log capital stock, log employment, firm-level innovation value, and industry-level innovation value), after standardizing all variables within each year and industry.  

\begin{table}[H]
  \caption{Correlations of PageRank Centrality and Log Changes (standardized within year and industry) with Firm Characteristics.}
  \label{tab:pagerank_corrs}
  \centering
  \begin{tabular}{lrrr}
    \toprule
    Variable & $\text{Centrality}^{PR}_{f,t}$ & $\Delta \text{Centrality}^{PR}_{f,t}$ (1-year) & $\Delta \text{Centrality}^{PR}_{f,t}$ (2-year) \\
    \midrule
    Log of profit   & 0.497*** & 0.002 & 0.007 \\
    Log of capital stock   & 0.461*** & -0.007 & -0.004 \\
    Log of employment   & 0.483*** & -0.002 & 0.002 \\
    Innovation value (firm)     & 0.404*** & 0.027*** & 0.046*** \\
    Innovation value (industry)    & -0.004 & 0.001 & 0.006 \\
    sale   & 0.378*** & -0.007 & -0.009 \\
    cost of goods sold   & 0.365*** & -0.006 & -0.008 \\
    \bottomrule
    \textit{Note:}&\multicolumn{3}{r}{  { 
  \(^{***}\) \(p<0.01\), \(^{**}\) \(0.01\le p<0.05\), \(^*\) \(0.05\le p<0.10\).}}
  \end{tabular}
  \vspace{0.5ex}
\end{table}

Table \ref{tab:pagerank_corrs} shows that innovation centrality is strongly and significantly correlated with firm size and success measures---profits, capital stock, employment, and firm-level innovation value. Larger, more profitable firms producing high-value innovations tend to occupy more central positions in the innovation network.  

By contrast, changes in innovation centrality (1- and 2-year growth) are only significantly associated with firm-level innovation value. This finding is consistent with expectations: firms generating high-value innovations in the current year are more likely to experience growth in innovation centrality relative to prior years. Importantly, this also highlights the value of focusing on \emph{growth} in innovation centrality in our econometric analyses---its variation is much less likely to be mechanically confounded by firm size or other static firm characteristics.  

Finally, as we show in Tables \ref{tab:profit_growth} and \ref{tab:profit_growth_robust}, growth in innovation centrality predicts medium- to long-run profit growth even after controlling for the focal firm’s innovation value. This indicates that while centrality and innovation value are significantly correlated, our centrality measure captures additional information about a firm’s profit growth potential that is not subsumed by innovation value alone.  

\subsubsection{Mechanism}
\label{app:mechanism}

Firms whose innovation centrality grows rapidly experience faster profitability growth because increases in centrality reflect a shift in the \emph{structural relevance} of their technologies, not merely the standalone value of their current inventions. When a firm’s patents become more central in the technology network, they are more likely to serve as reference points, complements, or standards for other firms’ innovations. This position generates diffusion leverage, expands complementarities across products and markets, and strengthens bargaining power in alliances and supply chains. Importantly, these effects boost profits without requiring a proportional increase in physical assets. As a result, profits rise faster relative to total assets, producing sustained increases in profitability ratios.  

To test this mechanism, we estimate regressions of profitability growth, defined as $\frac{\text{sale} - \text{cogs}}{\text{at}}$, on innovation centrality growth using a modified version of equation~\ref{eq:main_reg}, where the left-hand side is profitability growth rather than profit growth. The key regressor is the change in the focal firm’s innovation centrality, and we control for the contemporaneous innovation value of the firm’s innovations.  

First, we find that the focal firm's current innovation value is significantly positively associated with contemporaneous profitability. In addition, the results in Table \ref{tab:margin_growth} show innovation values are negatively associated with subsequent profitability growth, reflecting that highly innovative firms already earning high margins tend to grow more slowly in profitability. By contrast, increases in innovation centrality are significantly positively associated with profitability growth, and the association unfolds gradually over the medium to long run: centrality growth does not coincide with sharp concurrent jumps in profitability, but instead predicts persistent gains in revenues and margins over the following years. Thus, innovation centrality captures the embedded capacity of a firm’s technologies to shape and benefit from the broader innovation ecosystem, making it a forward-looking predictor of profit growth that is distinct from, and complementary to, the contemporaneous value of patents.

\begin{table}[H] \centering
  \caption{The association between innovation centrality growth and profit margin growth.} 
  \label{tab:margin_growth} 
\begin{tabular}{@{\extracolsep{5pt}}lrrrrr
} 
\\[-1.8ex]\hline 
\hline \\[-1.8ex] 
 & \multicolumn{5}{c}{\textit{Dependent variable:}} \\ 
\cline{2-6} 
\\[-1.8ex]  & \multicolumn{1}{c}{$\Delta M_{t,t+1}$} & \multicolumn{1}{c}{$\Delta M_{t,t+2}$} & \multicolumn{1}{c}{$\Delta M_{t,t+3}$} & \multicolumn{1}{c}{$\Delta M_{t,t+4}$} & \multicolumn{1}{c}{$\Delta M_{t,t+5}$} \\
\\[-1.8ex]  & \multicolumn{1}{c}{(1)} & \multicolumn{1}{c}{(2)} & \multicolumn{1}{c}{(3)} & \multicolumn{1}{c}{(4)} & \multicolumn{1}{c}{(5)} \\
 \hline \\[-1.8ex] $\text{Centrality}^{PR}_{f,t}$ (1-year)  & $0.000$ & $0.000^{**}$ & $0.002^{**}$ & $0.002^{***}$ & $0.003^{***}$ \\ 
  & $(0.000)$ & $(0.000)$ & $(0.001)$ & $(0.001)$ & $(0.001)$ \\ 
  Firm-level innovation value & $-0.000$ & $-0.002^{**}$ & $-0.003^{***}$ & $-0.005^{***}$ & $-0.005^{***}$ \\ 
  & $(0.001)$ & $(0.001)$ & $(0.001)$ & $(0.002)$ & $(0.002)$ \\ 
  $\text{Centrality}^{PR}_{f,t}$ (2-year)  & $0.000$ & $0.000$ & $0.003^{***}$ & $0.004^{**}$ & $0.003^{**}$ \\ 
  & $(0.000)$ & $(0.000)$ & $(0.001)$ & $(0.002)$ & $(0.001)$ \\ 
  Firm-level innovation value & $-0.001$ & $-0.002^{**}$ & $-0.003^{***}$ & $-0.005^{***}$ & $-0.006^{***}$ \\ 
  & $(0.001)$ & $(0.001)$ & $(0.001)$ & $(0.002)$ & $(0.002)$ \\

 \hline \\[-1.8ex] 
Observations PR 1y& \multicolumn{1}{c}{21,761} & \multicolumn{1}{c}{20,872} & \multicolumn{1}{c}{17,146} & \multicolumn{1}{c}{15,559} & \multicolumn{1}{c}{14,157} \\ 
R$^{2}$: PR 1y& \multicolumn{1}{c}{0.003} & \multicolumn{1}{c}{0.007} & \multicolumn{1}{c}{0.034} & \multicolumn{1}{c}{0.036} & \multicolumn{1}{c}{0.038} \\ 

Observations PR 2y & \multicolumn{1}{c}{20,988} & \multicolumn{1}{c}{20,312} & \multicolumn{1}{c}{16,600} & \multicolumn{1}{c}{15,102} & \multicolumn{1}{c}{13,732} \\ 
R$^{2}$: PR 2y& \multicolumn{1}{c}{0.006} & \multicolumn{1}{c}{0.006} & \multicolumn{1}{c}{0.038} & \multicolumn{1}{c}{0.039} & \multicolumn{1}{c}{0.041} \\  
\hline 
\hline \\[-1.8ex] 
\textit{Note:}  & \multicolumn{5}{r}{$^{*}$p$<$0.1; $^{**}$p$<$0.05; $^{***}$p$<$0.01} \\ 
\end{tabular} 
\end{table}

\end{document}